\documentclass[11pt]{article}

\usepackage[utf8]{inputenc}
\usepackage{amsmath,amssymb}
\usepackage{amsfonts}
\usepackage{booktabs}
\usepackage{siunitx}
\usepackage{algpseudocode}
\usepackage{algorithm}
\usepackage{enumitem}
\usepackage[margin=1in]{geometry}
\usepackage[authoryear,round]{natbib} 
\setcitestyle{aysep={}} 
\usepackage{graphicx}
\usepackage{soul}
\usepackage{setspace}
\usepackage{listings}
\usepackage{color,soul}
\usepackage{bbm}
\usepackage{authblk}
\usepackage{multirow}

\usepackage[svgnames]{xcolor}
\usepackage{listings}
\definecolor{backcolour}{rgb}{0.95,0.95,0.92}
\usepackage{tikz-cd}
\usetikzlibrary{positioning, arrows.meta}
\usepackage[most]{tcolorbox}

\usepackage{subcaption}

\newcommand{\vect}[1]{\textbf{\textit{#1}}}

\newcommand{\E}{\mathrm{E}}

\DeclareMathOperator*{\argmin}{arg\,min}
\newtheorem{theorem}{Theorem}

\newtheorem{proposition}{Proposition}
\usepackage{mathtools}

\newlist{condenum}{enumerate}{1} 
\setlist[condenum]{label=\bfseries Condition \arabic*., ref=\arabic*, wide}

\newlist{propnum}{enumerate}{1} 
\setlist[propnum]{label=\bfseries Proposition \arabic*., ref=\arabic*, wide}

\title{Title}
\title{Bad estimation, good prediction: the Lasso in dense regimes}
\author[1*]{Andrea Bratsberg}
\author[1]{Magne Thoresen}
\author[2]{Jelle J. Goeman}
\affil[1]{Oslo Centre for Biostatistics and
Epidemiology, Department of Biostatistics,
University of Oslo}
\affil[2]{Department of Biomedical Data Sciences,
Leiden University Medical Center}
\affil[*]{\textit{email: a.m.bratsberg@medisin.uio.no}}

\date{}

\begin{document}
\maketitle

\begin{abstract}
For high-dimensional omics data, sparsity-inducing regularization methods such as the Lasso are widely used and often yield strong predictive performance, even in settings when the assumption of sparsity is likely violated. 
We demonstrate that under a specific dense model, namely the high-dimensional joint latent variable model, the Lasso produces sparse prediction rules with favorable prediction error bounds, even when the underlying regression coefficient vector is not sparse at all. We further argue that this model better represents many types of omics data than sparse linear regression models. 
We prove that the prediction bound under this model in fact decreases with increasing number of predictors, and confirm this through simulation examples. These results highlight the need for caution when interpreting sparse prediction rules, as strong prediction accuracy of a sparse prediction rule may not imply underlying biological significance of the individual predictors.
\end{abstract}

\textbf{Keywords:} high-dimensional omics; sparsity; latent variable models; multicollinearity; eigenvalue decay.

\section{Introduction}
\label{seq:motivation}

Technological developments in the last few decades have improved our ability to measure vast amounts of data across various research fields, including omics, econometrics, and finance. In omics research (transcriptomics, in particular), microarrays and next-generation sequencing  has made it possible to quickly and cheaply measure gene expression from thousands of genes for a number of subjects, simultaneously. If properly analyzed, these data could improve our understanding of the human genome, which in turn can potentially enhance current research on personal treatment or precision medicine. A common objective in the development of new biomarkers, for example, is to identify subsets of genes or molecular networks that are predictive of a phenotype, such as disease status or prognosis \citep{segal2003regression}. 

To this end, sparse methods for high-dimensional regression models have become immensely popular due to their ability to produce parsimonious and interpretable solutions, even when the number of predictors far exceeds the number of independent samples. These methods rely on the assumption that only a small subset of predictors are truly relevant for the outcome of interest, while the rest are noise, and aim at selecting these non-zero effects (i.e., variable selection). One of the most widely used variable selection methods is the Lasso \citep{tibshirani1996lasso}, which favors solutions with small $\ell_1$-norm, resulting in a sparse estimated regression coefficient vector. Lasso and other sparsity-inducing methods often show good prediction performance across a wide range of settings, even when their underlying assumptions are likely violated \citep{ogutu2012genomic, wang2019precision, abraham2013performance, ajana2019benefits, greenshtein2004persistence, waldron2011optimized, zhao2010development}. Prediction is fundamentally an easier task than estimation, and it is well known that a good predictor does not necessarily provide an accurate estimate of the underlying model. Still, we believe there is a need to clarify this more explicitly for high-dimensional omics data, as sparse solutions are in practice often interpreted, either explicitly or implicitly, as being biologically meaningful. We argue in the following that the assumption of sparsity is often unrealistic for omics data and that researchers should be cautious when interpreting a good sparse prediction rule. Similar questions have been raised in other fields, such as econometrics \citep{giannone2021economic}.

\subsection{The sparsity assumption for omics data}
\label{sec:sparsityassumtionIntro}

First, omics data, such as gene expression measurements, are typically highly correlated. Gene expression data are the results of the
combinatorial effects from many highly connected biological pathways \citep{sun2004genomic}, resulting in potentially long-range correlations and strong local correlations. Additionally, the measured variables are inherently noisy, due to for example technical noise or biological variation, and even the noise may be correlated \citep{segura2019predictive}. Such highly correlated designs may contradict the assumption of a sparse underlying multiple regression model. If, for example, there is only one variable in a group of correlated variables that is directly related to the outcome, the remaining variables may partially explain its residual variation, leading to a dense representation in the true model. Thus, if many or all variables are highly correlated, then the resulting (multiple) regression coefficient will not be sparse \citep[Ch. 15]{klein2014handbook}. Empirical evidence also appears to support this issue; \citet{ein2005outcome} showed that for a single breast cancer survival data set, there are several different sets of 70 genes that are equally predictive of survival, undermining the notion of an underlying unique smaller subset of "driving genes". Some genomic data may coincide with the assumption of sparsity, such as single nucleotide polymorphisms (SNPs) \citep{goeman2020comments}. However, this is rather the exception than the rule. 

Second, sparse methods such as the Lasso rely on fairly strong assumptions to achieve model selection consistency, that is, the ability to recover the true sparse model when one exists. For example, the design matrix needs to fulfill the so-called irrepresentability condition \citep{zhao2006model}, which essentially says that the non-influential (or "non-active") variables cannot depend too strongly on the influential ("active") variables. When the number of variables exceeds the number of independent samples, the columns of the design matrix are necessarily linear combinations of each other. Thus, when the number of variables is extremely large compared to the number of independent samples, this condition is generally difficult to satisfy. In particular, \citet{wang2019precision} demonstrated that for three types of real omics data (gene expression, methylation and miRNA) and three cancer types, the irrepresentability condition is often violated even when the true model is sparse. Another key assumption for most sparse methods is that the truly influential variables must have large effects on the outcome of interest, while the remaining variables have negligible effects \citep{van2018tight}. This assumption is also often violated for genomic data, particularly for complex diseases, where correlations between each variable and the outcome seem to be much more distributed across the whole genome \citep{boyle2017expanded, goeman2020comments, ein2005outcome}. Even when there is strong evidence for the existence of "core genes" that directly affect a disease, and that can in principle be interpreted, these are affected by an abundance of tiny influences from other genes \citep{boyle2017expanded}. Consequently, such "core genes" may explain only a small part of the risk of a disease, even though they have a significant role in its development \citep{ball2023life, boyle2017expanded}. 

Given the noisy and highly correlated nature of omics data, the assumption that a unique subset of variables drives the outcome is often unrealistic. Even in settings where this assumption is plausible, the stringent conditions for sparse methods to select this set are rarely met in real omics data. While there exist many extensions to the Lasso, such as Elastic net \citep{zou2005elasticnet} that deals with the issue of collinearity by forcing a grouping structure to the coefficient estimates, the fundamental issue of sparsity still remains. Sparse methods are not stable, whereas stable solutions are not sparse \citep{xu2011sparse}. Efforts to stabilize the solution by aggregating the results from multiple models (ensemble methods) still lead to similar conclusions in high-dimensional omics settings; different subsets of variables with almost no overlap give comparable prediction performance \citep{pes2017exploiting}. 

\subsection{An alternative model}
\label{sec:introModel}

To better capture the complex interactions between genes and their effects phenotypes (e.g. disease risk), \citet[p. 1184]{boyle2017expanded} proposed that a more realistic model is "that disease risk is largely driven by genes with no direct relevance to disease and is propagated through regulatory networks to a much smaller number of core genes with direct effects". Motivated by this perspective, we propose the use of a joint latent variable model to represent the dependencies among observed variables and their relationship with the outcome \citep[chap. 6]{goemanthesis}. The latent variable model is a common approach to characterize dependency structures among features \citep{fan2024latent}, and it allows for the inclusion of random errors in the observed features such as measurement error, which is very common in omics data. These models assume that both the observed predictors (e.g., gene expression) and the outcome are driven by a common set of unobserved latent variables. These may be the shared metabolic pathways or regulatory mechanisms \citep{carvalho2008high, leek2011asymptotic}. In such models, any linear predictor based on the high-dimensional set of observed predictors is far from sparse. Yet, as we will illustrate, both through simulations and theoretical results, Lasso can still yield accurate predictions even when the underlying model is dense, and in particular as the number of predictors diverges. Our results align with the results of \citet{greenshtein2004persistence}, who showed that there is asymptotically no harm in introducing many more explanatory variables when predicting via the Lasso, and that "Occam's razor does not seem relevant for prediction".
Sparse solutions, such as those produced by Lasso, should thus be viewed as practical prediction tools rather than representations of the underlying biological reality.

\subsection{Illustrative example}
To illustrate our point, we revisit a simple (low-dimensional) example from the original Elastic net article \citep{zou2005elasticnet}. The Elastic net penalizes both large $\ell_1$-norm and $\ell_2$-norm solutions. This way, it decorrelates the variables (due to the $\ell_2$ penalty) and can be seen as a stabilizing version of the Lasso. In the example, they consider two latent variables $Z_1$ and $Z_2$ which are independent $\mathcal{U}(0, 20)$. The response $y$ is generated as
$\mathcal{N}(\boldsymbol{\beta}^T\mathbf{z},1)$, where $\mathbf{z} = (Z_1,Z_2)^T$ and $\boldsymbol{\beta} = (1,0.1)^T$. The observed variables are generated as $\mathbf{x} = A^T\mathbf{z} + \mathbf{e}$, where $\mathbf{e}\in \mathcal{N}(0,\frac{1}{16}I)$ and 
\begin{align*}
    A = \begin{pmatrix}
        1 & -1 & 1 & 0 & 0 &0 \\
        0 & 0 & 0 & 1 &-1&1
    \end{pmatrix}
\end{align*}
Thus, the first three variables form a group whose underlying factor is $Z_1$, and the remaining variables 
form a second group whose underlying factor is $Z_2$. The within-group correlations are almost 1 and the between-group correlations are almost 0. From Figure 5 in \cite{zou2005elasticnet}, it is clear that the Elastic net can be thought of as a stabilized version of the Lasso, with the latter showing a highly unstable variable selection.

But what about the prediction error? The best linear unbiased predictor of $y$ using only $\mathbf{x}$ is $\mathbf{x}^T\boldsymbol{\gamma}_0$, where $\boldsymbol{\gamma}_0 = (A^TA+\frac{1}{16}I)^{-1}A^T\boldsymbol{\beta}$. We may think of this as the true underlying regression coefficient. We now rerun the exact same example and instead record the out-of-sample prediction error $\|y_*-\mathbf{x}^{T}_*\hat{\boldsymbol{\gamma}}\|_2^2$ for a new data point $(\mathbf{x}^{T}_*,y_*)$, together with the $\ell_2$-estimation error $\|\boldsymbol{\gamma}_0-\hat{\boldsymbol{\gamma}}\|_2^2$, where $\hat{\boldsymbol{\gamma}}$ is either the Lasso or Elastic net estimate. We repeat the simulation 100 times and report the average, as a function of $s =  \|\hat{\boldsymbol{\gamma}}\|_1/\|\hat{\boldsymbol{\gamma}}_n\|_1$, where $\hat{\boldsymbol{\gamma}}_n$ is the full ordinary least squares estimate.

\begin{figure}[H]
\centering
\includegraphics[width=0.9\linewidth]{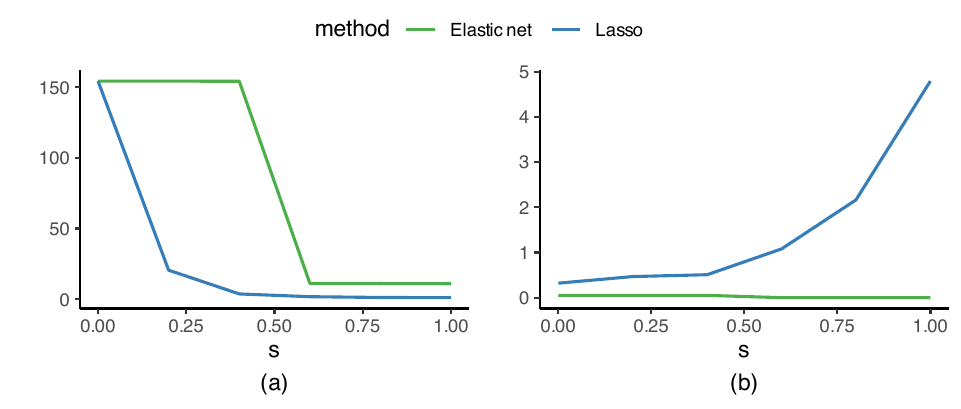}
  \caption{(a) Out-of-sample prediction error  $\|y_*-\mathbf{x}^{T}_*\hat{\boldsymbol{\gamma}}\|_2^2$, and (b) estimation error $\|\boldsymbol{\gamma}_0-\hat{\boldsymbol{\gamma}}\|_2^2$ for Lasso vs. Elastic net.} 
\label{fig:fig1}
\end{figure}

The display to the right in Figure \ref{fig:fig1} shows that the Elastic net gives a much more correct estimate of the underlying regression coefficient in terms of the $\ell_2$-estimation error. However, when we consider the prediction error (left display), Lasso clearly outperforms the Elastic net. By focusing on the results at, e.g., $s = 0.75$, we can draw two key conclusions: $(1)$ a good sparse estimator (the Elastic net) is not necessarily a good sparse predictor, and $(2)$ a good sparse predictor (the Lasso) is not necessarily a good sparse estimator.

\subsection{Outline of the paper}

The rest of this work is organized as follows. In Section \ref{sec:latentvariablemodel} we introduce and define the high-dimensional joint latent variable model, allowing the number of observed variables $p$ to approach infinity. In Section \ref{sec:lassoprediction}, we build on the theory for Lasso prediction for correlated designs and prove that, under the latent variable model, the in-sample prediction error exhibits favorable bounds for increasing $p$. Section \ref{sec:numericalstudies} presents simulation examples that explore how the prediction (and estimation) errors are influenced by $p$ and the correlation structure of the random error in the observed variables. 

\section{Joint latent variable model}
\label{sec:latentvariablemodel}

Assume we observe $p$ predictors $\mathbf{x} = (x_1,...,x_p)^T$ and response $y$, and that both are driven by a set of latent variables $\mathbf{f}=(f_1,...,f_m)^T$:
\begin{center}
\begin{tikzcd}
& & & \\
x_1,...,x_p & & y\\
& \arrow{lu} f_1,...,f_m \arrow{ru}& \\
\end{tikzcd}  
\end{center}

and that $m \ll p$. The variables $f_1,...,f_m$ can for example be shared biological pathways. In addition, the predictor variables are subject to (possibly correlated) random error, which consists of uncontrollable sources of variability, including sampling and technical measurement error. Specifically, we assume the following model:
\begin{align}
    y &= \boldsymbol{\beta}^T\mathbf{f}+\epsilon\nonumber  \\
    \mathbf{x} &= A^T\mathbf{f}+\mathbf{e},
    \label{EQ:latentvariablemodel}
\end{align}
where we have implicitly assumed without loss of generality that the marginal means of $y$ and $\mathbf{x}$ are zero. The parameters $\boldsymbol{\beta}$ and $A$ are an $m-$vector and an $m \times p$ matrix of loadings. We assume that $\epsilon$ and $\mathbf{e}$ are uncorrelated and that they have mean zero and variance-covariance $\sigma^2$ and $\Psi$, respectively. The joint vector $\mathbf{z} = (y,\mathbf{x}^T)^T$ then has mean zero and covariance matrix
\begin{align*}
    \Sigma_z = \begin{pmatrix}\boldsymbol{\beta}^T\boldsymbol{\beta}+\sigma^2 & \boldsymbol{\beta}^TA \\
A^T\boldsymbol{\beta} & A^TA+\Psi\end{pmatrix}.
\end{align*}

If we assume normality of $\mathbf{z}$, we have that 
\begin{align*}
    \E[y| \mathbf{x}] &= \boldsymbol{\gamma}_0^T\mathbf{x},  \quad \text{ where } \boldsymbol{\gamma}_0 = (A^TA+\Psi)^{-1}A^T\boldsymbol{\beta}.
\end{align*}
However, if normality is not assumed,$\boldsymbol{\gamma}_0^T\mathbf{x} $ is still the best linear unbiasedprediction of $y$ given $\mathbf{x}$.
We may therefore predict the mean of $y$ from $\vect{x}$ by
\begin{align*}
     \E[y| \mathbf{x}] = \boldsymbol{\gamma}_0^T\mathbf{x}.
\end{align*}

Under this high-dimensional latent variable model, $\boldsymbol{\gamma}_0$ is potentially very dense in the sense that the number of non-zero entries may be as large as $p$. 
The term $A^T\beta$ captures the direct correlation between the predictors and the outcome $y$ and the non-zero elements may correspond to the group of "core genes" (see Section \ref{sec:introModel}), while it is zero for the rest of the predictors. 

To allow for the possibility of diverging $p$, we need some additional assumptions to make the model well-defined. Assume that when $p$ grows, new columns are added to $A$, and rows and columns are added to $\Psi$, and impose the following restrictions on the matrices, similarly to that of \citet[Ch. 6]{goemanthesis}:
\begin{enumerate}
    \item[A.1] There are constants $0 < k \leq K < \infty$ such that all eigenvalues of $\Psi$ are between $k $ and $K$ for all $p$.
    \item[A.2] The limit $\lim_{p \to \infty} \frac{1}{p} AA^T \in \mathbb{R}^{m\times m}$ exists and is of full rank $m$. 
\end{enumerate}

We may note a few implications of A.1 and A.2. First, they neatly separate the covariance matrix into a structural part, $A^TA$, and a noise part $\Psi$. Thus, $\mathbf{x}$ can be considered coming from a spiked covariance model \citep{johnstone2001distribution}, i.e. that the ordered eigenvalues $\omega_1, ... ,\omega_p$ of $\E[\mathbf{x}\mathbf{x}^T] = A^TA+\Psi$ are distributed as follows:
\begin{align*}
    \omega_1 > \omega_2 > \cdots > \omega_{m}  \gg   \omega_{m+1} \geq \cdots \geq \omega_p > 0,
\end{align*}
where the "non-spiked" eigenvalues $\omega_{m+1},...,\omega_p$ are bounded due to A.1, and the spiked eigenvalues $\omega_1,...,\omega_m = \mathcal{O}(p)$ due to A.2. Second, A.2 ensures that there is a non-vanishing proportion of non-zero entries in $A$ as the dimension $p$ grows, which is referred to as the \textit{pervasiveness} assumption in the latent factor model literature \citep{fan2013large}. This means that when the number of variables $p$ is very large, we have abundant information about the underlying factor $\mathbf{f}$ in $\mathbf{x}$. 

While the number of non-zero entries of $\boldsymbol{\gamma}_0$ grows with $p$, the conditions A.1 and A.2 in fact ensure that $\|\boldsymbol{\gamma}_0\|_1$ is approximately sparse in one specific way; the $\ell_1$-norm stays bounded for $p \to \infty$, as shown in the following proposition, with proof given in the Appendix. Note, however, that $\|\boldsymbol{\gamma}_0 \|_1$ may be large if $K \gg k$.
\begin{proposition}
Under A.1 and A.2, 
\begin{align}
\|\boldsymbol{\gamma}_0\|_1 = \mathcal{O}(1).
\label{eq:bound_l1_norm}
\end{align}
\end{proposition}
This proposition indicates that while the assumption of sparsity (in the $\ell_0$ sense) implies a small $\ell_1$-norm, the converse is not necessarily true. The joint latent variable model \eqref{EQ:latentvariablemodel} is thus an example of a situation with small $\ell_1$-norm but no sparsity.

As argued in Section \ref{sec:introModel}, this latent variable model is suitable for modelling the types of data that have become common in omics research. We now turn to show that under the latent variable model, a sparse Lasso solution gives favorable prediction bounds, even when $p \to \infty$, if the regularization parameter is chosen accordingly.

\section{Lasso prediction under the joint latent variable model}
\label{sec:lassoprediction}

Suppose we have $n$ observations from the model \eqref{EQ:latentvariablemodel}, giving us the design matrix $X \in \mathbb{R}^{n \times p}$ and the vector of observed outcomes $\mathbf{y} \in \mathbb{R}^{n}$. The Lasso estimator \cite{tibshirani1996lasso} is defined as 
\begin{align}
    \hat{\boldsymbol{\gamma}} = \argmin_{\boldsymbol{\gamma} \in \mathbb{R}^p}\left\{\frac{1}{n}\|\mathbf{y}-X\boldsymbol{\gamma}\|_2^2+\lambda\|\boldsymbol{\gamma}\|_1\right\},
    \label{EQ:penalizedformLasso}
\end{align}
where $\lambda \geq 0$ is a regularization parameter that penalizes large values of $\|\boldsymbol{\gamma}\|_1$ and typically forces many of the entries of $\boldsymbol{\gamma}$ to be exactly zero.


For large $p$, any sparse approximation $\hat{\boldsymbol{\gamma}}$ of $\boldsymbol{\gamma}_0= (A^TA+\Psi)^{-1}A^T\boldsymbol{\beta}$ may quickly become very different from $\boldsymbol{\gamma}_0$ in terms of the estimation error $\|\hat{\boldsymbol{\gamma}}-\boldsymbol{\gamma}_0\|_2^2$ since the number of non-zero entries of $\boldsymbol{\gamma}_0$ grows linearly with $p$. However, for prediction accuracy, it is sufficient that $\|\boldsymbol{\gamma}_0\|_1$ remains bounded, as shown below.

We use the following measure of prediction performance:
\begin{align*}
    \text{MSE}(\hat{\boldsymbol{\gamma}}) &= \frac{1}{n}\|X\boldsymbol{\gamma}_0-X\hat{\boldsymbol{\gamma}}\|_2^2  = (\boldsymbol{\gamma}_0-\hat{\boldsymbol{\gamma}})^T\hat{\Sigma}(\boldsymbol{\gamma}_0-\hat{\boldsymbol{\gamma}})
\end{align*}
where $\hat{\Sigma} = X^TX/n$. Note that under the normality assumption for $(y, \mathbf{x}^T)$, $\text{MSE}(\hat{\boldsymbol{\gamma}})$ differs from the prediction error for fixed designs, $\frac{1}{n}\|\mathbf{y}-X\hat{\boldsymbol{\gamma}}\|_2^2$, by an amount $\sigma_{|x}^2$, defined as:  $$\sigma_{|x}^2 := \boldsymbol{\beta}^T\left(I-(A\Psi^{-1}A^T+I)A\Psi^{-1}A^T\right)\boldsymbol{\beta} + \sigma^2.$$
When $p$ is large, the identity matrix becomes negligible compared to $A\Psi^{-1}A^T$, and $\sigma_{|x}^2 \approx \sigma^2$ \citep{goemanthesis}. Hence, $\text{MSE}(\hat{\boldsymbol{\gamma}})$ adequately reflects the excess prediction risk for the fixed design case.

The simplest bound for Lasso prediction error, valid for an arbitrary design matrix $X$, is given by:
\begin{align}
         \text{MSE}(\hat{\boldsymbol{\gamma}}) = \frac{1}{n}\|X\boldsymbol{\gamma}_0-X\hat{\boldsymbol{\gamma}}\|_2^2 \leq \frac{2}{n}\lambda \|\boldsymbol{\gamma}_0\|_1.
    \label{EQ:simpleboundLasso}
\end{align}
on the set $\mathcal{T} = \{\sup_{\boldsymbol{\gamma}}2\sigma\|\epsilon^TX\boldsymbol{\gamma}\|/\|\boldsymbol{\gamma}\|_1\leq \lambda\}$ \cite[Eq.(2)]{hebiri2012correlations}. By Proposition 1, this bound may be favorable and it holds for any value of $\lambda$. However, some $\lambda$ make the probability of being on the set $\mathcal{T}$ higher than others. Usually, $\lambda$ is set to be proportional to $\sigma\sqrt{n\log(p)}$. 
However, this choice of $\lambda$ is not optimal if the columns of $X$ are highly correlated, as in the case of model \eqref{EQ:latentvariablemodel} when $p$ is large; the optimal choice of $\lambda$ is typically smaller for highly correlated designs than for orthogonal or weakly correlated designs \citep{van2013lasso, hebiri2012correlations}. Still, for nearly collinear designs, \citet{dalalyan2017prediction} shows that even the universal
tuning parameter $\lambda \propto \sigma\sqrt{n\log(p)}$ can yield favorable prediction bounds for the Lasso.

Under the joint latent variable model \eqref{EQ:latentvariablemodel}, a faster prediction error bound than \eqref{EQ:simpleboundLasso} is achievable for a properly chosen penalty parameter $\lambda$. To establish this, we utilize results from \citet{van2013lasso}, which addresses highly correlated designs. They consider a design matrix to be highly correlated if the rate at which the eigenvalues of $\hat{\Sigma}:= X^TX/n$ decay is sufficiently fast. The structural assumption of spectral decay has recently gained more attention for its agreement with real-life examples; see \citet{silin2022canonical} and references therein. For the model \eqref{EQ:latentvariablemodel}, condition A.2 (pervasiveness assumption) implies that the $m$ spiked eigenvalues $\omega_1, ..., \omega_m$ grow linearly with $p$, while the non-spiked eigenvalues $\omega_{m+1}, ..., \omega_p$ will increase in number but remain bounded (due to Condition A.1). Consequently, the ordered set of empirical eigenvalues $\tilde{\omega}_1,\tilde{\omega}_2, ... , \tilde{\omega}_p$ will indeed decay fast when $p$ grows. The rate of this decay in the limit $p \to \infty$ depends on the ratio $p/n$ and the magnitude of the leading population eigenvalues $\omega_1,...,\omega_m$ \citep{wang2017asymptotics, cai2020limiting}. See \citet{baik2006eigenvalues} for the (almost sure) relationship between $\tilde{\omega}_i$ and $\omega_i$ for the case where $\omega_{m+1},...,\omega_p = 1$ and $p, n \to \infty$. By adapting Lemma 6.1 and Corollary 5.2 from \cite{van2013lasso} to
our setting, we derive an improved bound, as stated in Theorem 1. The proof is given in the Appendix.

\begin{theorem}
Assume that A.1 and A.2 hold and $p  \geq n$. Then, there exsists positive constants $C$ and $c_0$ such that for any $t>0$, $p \geq c_0$, and
\begin{align*}
    \lambda = \left[C\sigma^2\left(\frac{(m+c_0p^{-1})^{1/2}}{\sqrt{2}-1}+t\right)\right]^{4/3}\|\boldsymbol{\gamma}_0\|_1^{-1/3}
\end{align*}
the bound 
\begin{align}
    \frac{1}{n}\|X(\boldsymbol{\gamma}_0-\hat{\boldsymbol{\gamma}})\|_2^2 \leq \frac{21}{2n}\left[C\sigma^2\left(\frac{(m+c_0p^{-1})^{1/2}}{\sqrt{2}-1}+t\right)\right]^{4/3}\|\boldsymbol{\gamma}_0\|_1^{2/3}
    \label{EQ:predictionbound}
\end{align}
holds with probability at least $1-\exp[-t^2]$.
\end{theorem}

\begin{description}
   \item[Remark 1] We did not attempt to optimize the constants in \eqref{EQ:predictionbound} and the bound may not be tight. 
\item[Remark 2] 
The constant $c_0$ is largely determined by $\Psi$ and its eigenvalues (see the proof of Theorem 1 for details); in particular by the quantity
\begin{align*}
    \frac{1}{\tilde{\omega}_1}\sum_{j=m+1}^n\tilde{\omega}_j,
\end{align*}
which we will refer to as the "partial effective rank". This is related to the idea of effective rank found in the literature on high-dimensional
sample covariance estimators \citep{koltchinskii2017}. 
\end{description}

Under the latent variable model, the prediction bound \eqref{EQ:predictionbound} in fact decreases with increasing $p$, and increases with increasing $m$. This is in sharp contrast to the prediction error bounds for the high-dimensional linear model under the sparsity assumption, in which increasing the number of (noise) variables can only hurt the prediction accuracy \citep{flynn2017sensitivity}. However, as mentioned in Section \ref{sec:introModel}, our results align well with the conclusions drawn in \cite{greenshtein2004persistence}, that increasing the number of variables $p$ does not hurt the prediction accuracy of a sparse Lasso solution. In the subsequent sections, we complement these theoretical insights through numerical studies, illustrating how the prediction accuracy is affected by the dimensionality and the correlation structure of $X$.

\section{Numerical studies}
\label{sec:numericalstudies}

For data generation, we set the number of latent variables to $m = 3$. We let the number of non-zero entries of each row of $A_j$, $j \in \{1,..,3\}$ be to $0.2p$, i.e., they are pervasive factors. The entries of $A$ are fixed, but they are generated from $\mathcal{U}(-1,1)$. The latent variables $\mathbf{f} \sim \mathcal{N}(0,1)$ and the noise $\epsilon \sim \mathcal{N}(0,1)$. We fix $n = 100$, but vary the number of variables $p\in \{5, ..., 10000\}$. 
We consider four settings of $\Psi$:
\begin{enumerate}
    \item $\Psi_1$ is the $p \times p $ identity matrix,
    \item $\Psi_2 = \text{diag}(\sigma_1^2,...,\sigma_p^2)$, where a random subset of size $p/2$ are considered common variants (low variance) and the other $p/2$ are rare (high variance). The low variances are generated from $\mathcal{U}(0.01,0.1)$ and the high variances from $\mathcal{U}(0.5,2)$,
    \item $\Psi_3$ has block-diagonal structure, and Toeplitz correlation structure within blocks. The correlation coefficient within blocks is set to 0.8 and the between-blocks correlation is 0.1. In addition, the eigenvalues of $\Psi$ are forced to be between $0.1$ and $3$. The number of blocks is set to $\lceil \sqrt{p} \rceil$.
    \item $\Psi_4$ is a random non-diagonal matrix with eigenvalues between $0.1$ and $3$ for all $p$. 
\end{enumerate}
Simulation settings 1 and 2 correspond to uncorrelated error, while settings 3 and 4 have correlated error. To evaluate the out-of-sample prediction error,  we consider not only  $\text{MSE}(\hat{\boldsymbol{\gamma}})$, but also 
\begin{align*}
    \text{PE}(\hat{\boldsymbol{\gamma}}) &= \E[(\mathbf{x}^T\boldsymbol{\gamma}_0-\mathbf{x}^T\hat{\boldsymbol{\gamma}})^2] = (\boldsymbol{\gamma}_0-\hat{\boldsymbol{\gamma}})^T\Sigma(\boldsymbol{\gamma}_0-\hat{\boldsymbol{\gamma}}),
\end{align*}
where $\Sigma = A^TA+\Psi$. Similar to $\text{MSE}(\hat{\boldsymbol{\gamma}})$ in the fixed design setting, $\text{PE}(\hat{\boldsymbol{\gamma}})$ reflects the excess prediction risk in the random design case, particularly for large $p$. We include this additional measure of prediction error because, in high dimensions, the matrices $\hat{\Sigma}$ and $\Sigma$ may differ considerably, and it is thus of interest to see how this is reflected in the prediction performance of the Lasso. Since $\boldsymbol{\gamma}_0$ also changes with $p$, we report the \textit{relative} MSE and PE, i.e. $\text{MSE}(\hat{\boldsymbol{\gamma}})/\text{MSE}(0)$ and $\text{PE}(\hat{\boldsymbol{\gamma}})/\text{PE}(0)$.

Even when $p > n$, the Lasso cannot select more than a total of $n$ variables. Let $\hat{\boldsymbol{\gamma}}_{n}$ be the Lasso estimate for the maximum number of selected variables. For simpler comparisons, we report the results for $s \in \{0,1\}$, where $s = \|\hat{\boldsymbol{\gamma}}\|_1/\|\hat{\boldsymbol{\gamma}}_{n}\|_1$ instead of $\lambda$, but we note that there exists a $\lambda$ for any $s$ that gives the same solution and that larger values of $s$ correspond to smaller $\lambda$. We also consider the standardized estimation error $\|\hat{\boldsymbol{\gamma}}- \boldsymbol{\gamma}_0\|_2^2/\|\boldsymbol{\gamma}_0\|_2^2$. We repeated the simulation 100 times and reported the average. 

\begin{figure}[!htb]
\centering
\includegraphics[width=0.8\linewidth]{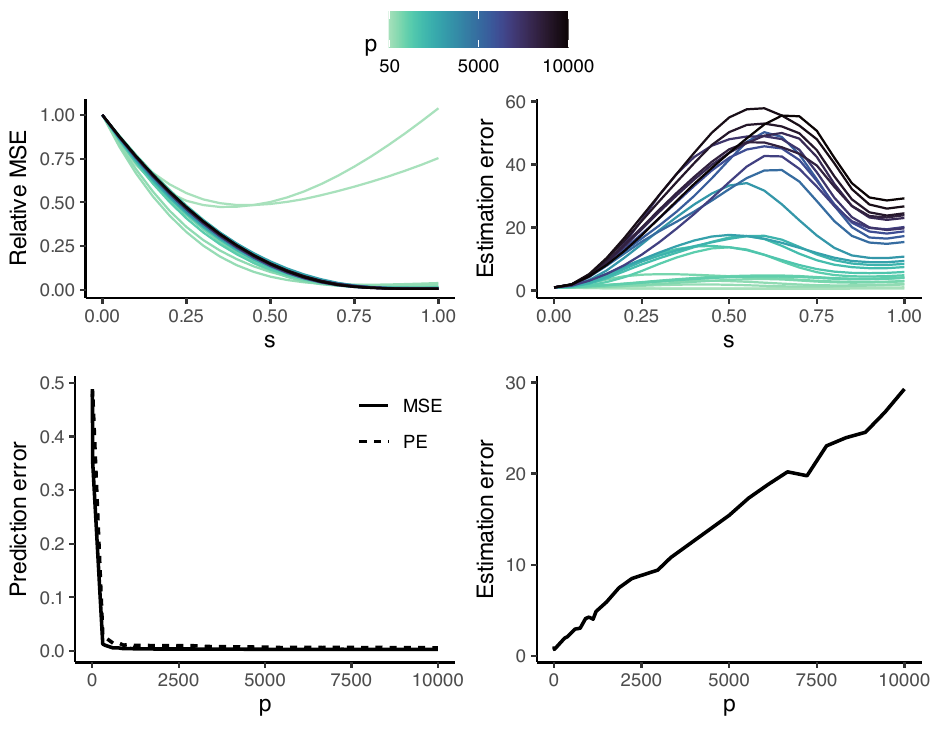}
\caption{Top display: the relative MSE (left) and standardized $\ell_1$ estimation error (right) for $s \in \{0,...,1\}$ for different number of variables $p$. Bottom display: the relative MSE and relative PE at the optimal choice of $s$ (left), and the $\ell_1$ estimation error for $s = 0.5$ (right). $\Psi = \Psi_3$.}
\label{fig:fig6}
\end{figure}

Figure \ref{fig:fig6} shows the results for the case $\Psi = \Psi_3$. Note that we do not include the results for $p = n = 100$ as this often results in a much worse prediction performance that will overshadow the other curves. This worsening of performance is associated with the "double descent" phenomenon, which has recently gained much attention in the machine learning literature \citep{belkin2019reconciling}.

From Figure \ref{fig:fig6}, it is clear that the prediction error of the Lasso indeed decreases for increasing number of variables $p$, while the estimation error naturally increases. Similar figures for the other cases can be found in Web Appendix A. The decay of the relative MSE as a function of $p$ (bottom left in Fig. \ref{fig:fig6}) is also nicely captured by the in-sample prediction bound \eqref{EQ:predictionbound}. The difference between the in-sample and out-of-sample prediction error in this case is minimal, showing that the predictions are stable with respect to new observations.

\begin{figure}[!htb]
\centering
\includegraphics[width=\linewidth]{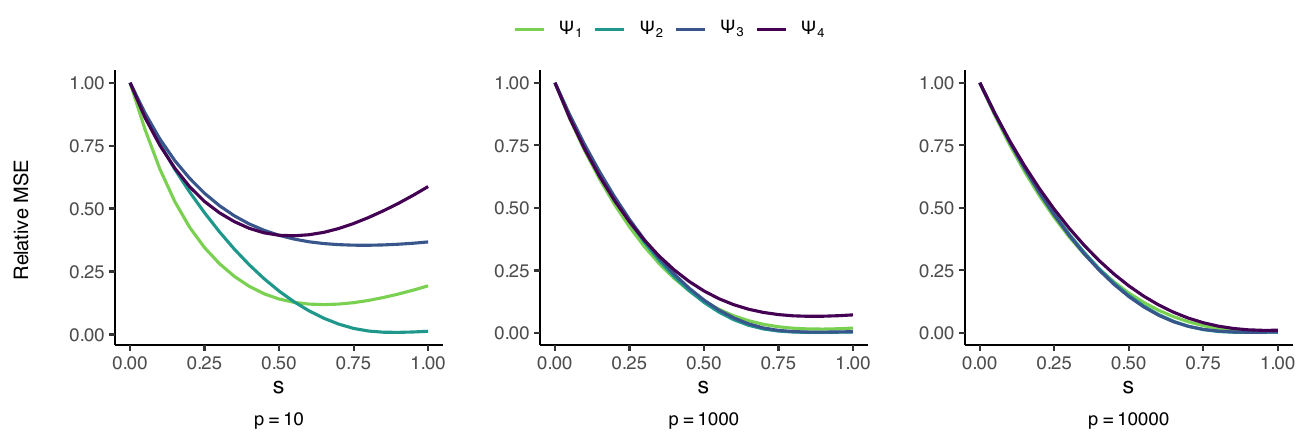}
\caption{The relative MSE for $p \in \{10, 1000, 10000\}$ for the four different examples of $\Psi$. The structure of $\Psi$ becomes less influential when the number of variables $p$ is large.}
\label{fig:comparisonsMSE}
\end{figure}

\begin{figure}[!htb]
\centering
\includegraphics[width=\linewidth]{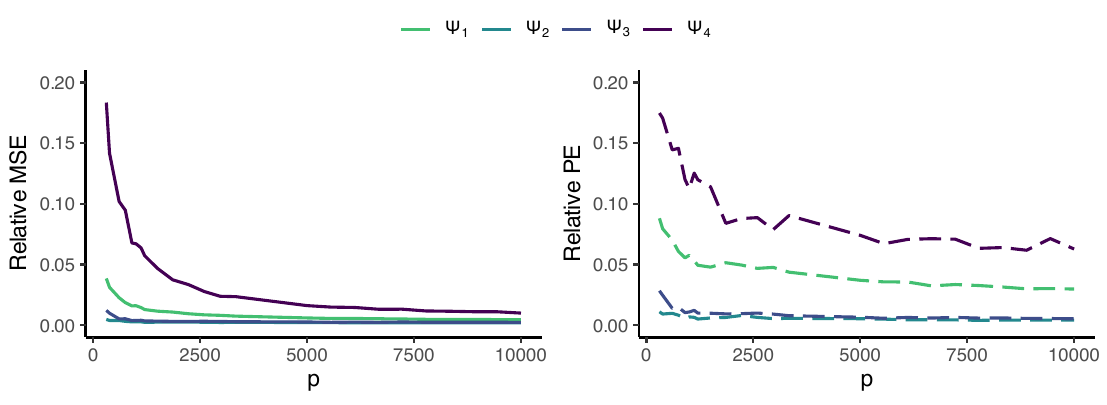}
\caption{The relative MSE (left) and PE (right) at the optimal choice of $s$ for the different choices of $\Psi$, as a function of $p > n$.}
\label{fig:comparisonsMSEandPE}
\end{figure}

Figure \ref{fig:comparisonsMSE} displays the relative MSE for different settings of $\Psi$, across three values of $p$.
The merging of the four curves as $p$ increases suggests that the structure of $\Psi$ becomes less influential as $p$ increases, which is due to the corresponding strengthening of the signal (under A.1 and A.2).

Figure \ref{fig:comparisonsMSEandPE} shows the prediction performance at the optimal value of $s \in \{0,1\}$ as a function of $p > n$ for $\Psi_1,...,\Psi_4$. As expected, the out-of-sample prediction error (PE) is consistently larger than the in-sample prediction error (MSE) across all structures of $\Psi$. However, PE shows a similar decay as $p$ increases. Furthermore, it is clear that both MSE and PE are worse for $\Psi_1$ and $\Psi_4$ compared to $\Psi_2$ and $\Psi_3$. This can indeed be related to the constant $c_0$ in \eqref{EQ:predictionbound}, as the partial effective ranks of the resulting design matrices corresponding to $\Psi_1$ and $\Psi_4$ are much larger than for $\Psi_2$ and $\Psi_3$ (see Fig. 4 in Web Appendix A). However, the constant $c_0$ determines only an upper bound and does not completely account for the ranking of the different $\Psi$ structures, which depends not only on the eigenvalues of $\Psi$ but also on its eigenvectors.

As additional exploration (Web Appendix B),
we show that under the latent variable model, there exist multiple disjoint subset that are have comparable predictive accuracy. This phenomenon is further illustrated using the real breast cancer dataset of \citet{vanVijver2002gene}.

\section{Discussion}

The aim of this work has been to highlight the importance of caution when interpreting sparse prediction rules, particularly in the context of high-dimensional omics data. While it is well understood that a strong predictive rule does not necessarily imply an accurate representation of the underlying truth, the use of sparse methods are often motivated by the fact that they are interpretable. For example, when genes, SNPs, or proteins are identified as potential novel biomarkers using sparse methods, they are frequently reported as "important variables" and implicitly regarded as potential targets for future treatment or intervention. It is, however, important to distinguish between two different goals: to find disease-related "master" genes, or to construct a prognostic tool. 

Nonetheless, an accurate sparse prediction rule can be extremely useful. For instance, in prognostic tests, a sparse biomarker panel, comprising only a few gene expressions or proteins, reduces the need for extensive measurements (e.g. the whole genome) for each patient \citep{rahnenfuhrer2023statistical}. Sparse predictors are thus economical and efficient, and makes the results more actionable for both clinicians and researchers \citep[Ch. 15]{klein2014handbook}. While a sparse prediction rule in the $p \gg n$ setting is not stable, it does not mean that the selected variables are not useful, as they are indeed very useful in terms of prediction; they are unstable as a prediction rule, but stable as a prediction. Furthermore, as $p \gg n$ and the true model is dense, as in latent variable models, there is no hope of estimating the regression coefficient $\gamma_0$ correctly. Hence, we might as well use a sparse predictor. This is the "bet on sparsity" principle, as coined by \citet{hastie_09_elements-of.statistical-learning}.

It is not surprising that the Lasso yields good prediction results in the present setting. As $p$ increases, so does the information about the underlying factors. As $p \gg n$, many columns of $X$ are near linear combinations of each other, and Lasso tends to select the optimal linear combinations in terms of MSE. An interesting further study would be to examine the variable selection properties of the Lasso in the limit $p \to \infty$, as the selected variables by the Lasso would perhaps resemble the directions chosen by partial least squares.

In this work, we consider the setting where $p$ diverges while $m$ remains fixed. When $m$ is very large, the prediction error bound \eqref{EQ:predictionbound} becomes correspondingly large. However, a large $m$ implies that the problem is not sparse in any sense of the word, and it cannot be expected that the Lasso would be an appropriate method. In certain situations, it is reasonable to assume that $m$ also grows with $p$, albeit at a slow rate. We believe that our theoretical results can be extended to this case,  but this is beyond the scope of this current work.

Our results confirm the virtue of regularization techniques when working with high-dimensional data to avoid overfitting and serve as an illustration of the conclusions drawn in \cite{greenshtein2004persistence}; under the present model, there is no harm in introducing many more explanatory variables than observations as long as some constraint (on the $\ell_1$-norm) is placed on the solution.  

\newpage

\bibliography{references}

\newpage

\appendix

\section{Proofs for Section \ref{sec:lassoprediction}}

\subsection{Proof of Proposition 1}

First note that $\boldsymbol{\gamma}_0 = (A^TA + \Psi)^{-1}A^T\boldsymbol{\beta} =  \Psi^{-1}A^T(A\Psi^{-1}A^T+I)^{-1}\boldsymbol{\beta}$ due to the Woodbury identity. We then have that
\begin{align*}
    \|\boldsymbol{\gamma}_0\|_1 = \|\Psi^{-1}A^T(A\Psi^{-1}A^T+I)^{-1}\boldsymbol{\beta}\|_1 \\
    \underset{(i)}{\leq} \sqrt{p}\|\Psi^{-1}A^T(A\Psi^{-1}A^T+I)^{-1}\boldsymbol{\beta}\|_2 \\\leq \sqrt{p}\|\Psi^{-1}\|_2\|A^T\|_2\|(A\Psi^{-1}A^T+I)^{-1}\|_2\|\beta\|_2,
\end{align*}
where $(i)$ is because $\|x\|_1 \leq \sqrt{d}\|x\|_2$ for any $d$-vector $x$. Now, $\|\Psi^{-1}\|_2 = 1/k$ due to A.1, $\|A^T\|_2 = \mathcal{O}(\sqrt{p})$ due to A.2 and $\|(A\Psi^{-1}A^T+I)^{-1}\|_2 = \mathcal{O}(K/p)$ and $\|\beta\|_2 = \mathcal{O}(1)$. Hence,
\begin{align*}
\|\boldsymbol{\gamma}_0\|_1 \leq \frac{\sqrt{p}}{k}\mathcal{O}(\sqrt{p})\mathcal{O}(Kp^{-1}) = \mathcal{O}(1).
\end{align*}
\hfill $\blacksquare$

\subsection{Proof of Theorem 1}

We follow the lines of the proof of Lemma 6.1 of \cite{van2013lasso} to show that Corollary 5.2 of \cite{van2013lasso} holds in our setting. 
Here, they assume that $\|X_j\|_2^2/n \leq 1$  for each column $X_j$, $j \in \{1,...,p\}$ of $X$. To fulfill this assumption in our case that allows for $p \to \infty$, we will consider a scaled version of $X$, $X' = \frac{1}{\sqrt{\tilde{\omega}_1}}X$, where $\sqrt{\tilde{\omega}_1}$ is the maximum singular value of $X$. Then, for any $p$, $\|X'_j\|_2^2/n\leq 1$. 
Let $\hat{\Sigma} = U\Omega U^T$ be the eigendecomposition of $\hat{\Sigma}$, where $U^TU=UU^T = I$ and $\Omega = \text{diag}(\tilde{\omega}_1,...,\tilde{\omega}_p)$, and since $p \geq n$, the last $n-p$ eigenvalues are equal to zero. The spiked sample eigenvalues $\tilde{\omega}_1,...,\tilde{\omega}_m = \mathcal{O}(p)$, while the remaining $n-m$ sample eigenvalues are bounded for any $p$ \cite{cai2020limiting}.
Denote by $\mathbf{u}_i$ the $i$th column of $U$ and note that $\sum_{i=1}^p (\mathbf{u}_i^T\boldsymbol{\gamma})^2 = \boldsymbol{\gamma}^TUU^T\boldsymbol{\gamma} = \|\boldsymbol{\gamma}\|_2^2\leq \|\boldsymbol{\gamma}\|_1 =1$.
Let $\mathcal{F} := \{X'\boldsymbol{\gamma}: \|\boldsymbol{\gamma}\|_1 = 1\}$ and consider any $f \in \mathcal{F}$ and note that since $\|\boldsymbol{\gamma}\|_1 = 1$,  $\|f\|_n \leq 1$, and  
\begin{align}
    \|f\|_n^2 = \frac{1}{n}\sum_{i=1}^n f_i = \frac{1}{\tilde{\omega}_1}\boldsymbol{\gamma}^T\hat{\Sigma}\boldsymbol{\gamma} = \sum_{i = 1}^p\frac{\tilde{\omega}_i}{\tilde{\omega}_1}(\mathbf{u}_i^T\boldsymbol{\gamma})^2 = \sum_{i = 1}^m \frac{\tilde{\omega}_i}{\tilde{\omega}_1}(\mathbf{u}_i^T\boldsymbol{\gamma})^2+\sum_{i = m+1}^n \frac{\tilde{\omega}_i}{\tilde{\omega}_1}(\mathbf{u}_i^T\boldsymbol{\gamma})^2 \nonumber\\ 
    \leq  \frac{\tilde{\omega}_1}{\tilde{\omega}_1}\sum_{i = 1}^m(\mathbf{u}_i^T\boldsymbol{\gamma})^2+\frac{\tilde{\omega}_{m+1}}{\tilde{\omega}_1}\sum_{i = m+1}^n (\mathbf{u}_i^T\boldsymbol{\gamma})^2 \leq 1 + \frac{c_0}{p} \leq 1,
    \label{EQ:sizeofF}
\end{align}
for some positive constant $c_0$, where the last inequality is due to the fact that $\tilde{\omega}_{m+1}/\tilde{\omega}_1 = \mathcal{O}(p^{-1})$ and $\sum_{i = 1}^p(\mathbf{u}_i^T\boldsymbol{\gamma})^2 \leq 1$. 

Define $f_m := P(f)$ as the projection operator that maps $f$ to a $m$-dimensional vector so that $\|f_m\|_n^2 = \sum_{i = 1}^m \frac{\tilde{\omega}_i}{\tilde{\omega}_1}(\mathbf{u}_i^T\boldsymbol{\gamma})^2$, and $\mathcal{F}_m := \{P(f): f \in \mathcal{F}\}$. Let $f_{m^c} = f-f_m$ be the part of $f$ orthogonal to $f_m$. Then,
\begin{align*}
\|f\|_n^2 = \|f_m\|_n^2 + \|f_{m^c}\|_n^2 \underset{(i)}{\leq} \|f_m\|_n^2 + \frac{c_0}{p} \implies \|f_{m^c}\|_n = \|f-f_m\|_n \leq \sqrt{\frac{c_0}{p}},
\end{align*} 
where $(i)$ is due to \eqref{EQ:sizeofF}.

Define $\delta := \sqrt{c_0p^{-1}}$. Since $p \geq c_0$, $0 < \delta \leq 1$. 
Let $\{x_1,...,x_N\} \subset \mathbb{R}^m$ be the minimal $\delta$-covering of $\mathcal{F}_m$ with corresponding cover number $N:=N(\delta, \mathcal{F}_m,\|\cdot\|_n)$, i.e., the minimal number of balls of radius $\delta$ with respect to $\|\cdot\|_n$ required to cover the set. Note that since $\|f_m\|_n \leq 1$, $\mathcal{F}_m$ is an $m$-ball with radius $1$, whose $\delta$-covering number is upper bounded by $(3/\delta)^m$ \cite[Lemma 14.27]{Bühlmann2011}.

Now, for each $f_m \in \mathcal{F}_m $, $\|f_m-x_i\|_n \leq \delta$ for some $x_i$,  $i \in \{1,...,N\}$ (by definition of $\delta$-covering). Let $\{\tilde{x}_i,...,\tilde{x}_N\} \in \mathbb{R}^n$ be the set consisting of each $x_i$ appended with $n-m$ zeros. Thus, for all $f \in \mathcal{F}$,
\begin{align*}
    \|f-\tilde{x}_i\|_n \leq \|f-f_m\|_n+\|f_m-x_i\|_n \leq 2\delta.
\end{align*}
Hence,  $\{\tilde{x}_i,...,\tilde{x}_N\}$ is a $2\delta$-covering number of $\mathcal{F}$ (not necessarily  minimal), and we get that
\begin{align*}
    N(2\delta, \mathcal{F}, \|\cdot\|_n) \leq N(\delta, \mathcal{F}_m, \|\cdot\|_n) \leq \left(\frac{3}{\delta}\right)^m \\
    \implies N(\delta, \mathcal{F}, \|\cdot\|_n) \leq \left(\frac{6}{\delta}\right)^m \\
    \implies \log(1+2N(\delta, \mathcal{F},\|\cdot\|) \leq \left(\frac{m+c_0p^{-1}}{\delta}\right),
\end{align*}
where $\delta = \sqrt{c_0p^{-1}}$. Thus, Corollary 5.2 \citet{van2013lasso} holds with $ A= m+c_0p^{-1}$ and $\alpha = 1/2$. This means that the bound \eqref{EQ:predictionbound} holds with probability at least $1-\exp[-t^2](1+2B)$, where 
\begin{align*}
    B:= \left(\exp\left[\frac{m+c_0p^{-1}}{4(\sqrt{2}-1)^2}-1\right]\right)^{-1}.
\end{align*}
$B$ becomes extremely small even for small $m$ and $c_0$, so we set $B \approx 0$. This completes the proof. 
\hfill $\blacksquare$

\newpage

\section{Supplemental simulation results}
\label{APP:moresimulations}

\begin{figure}[H]
\centering
\includegraphics[width=0.9\linewidth]{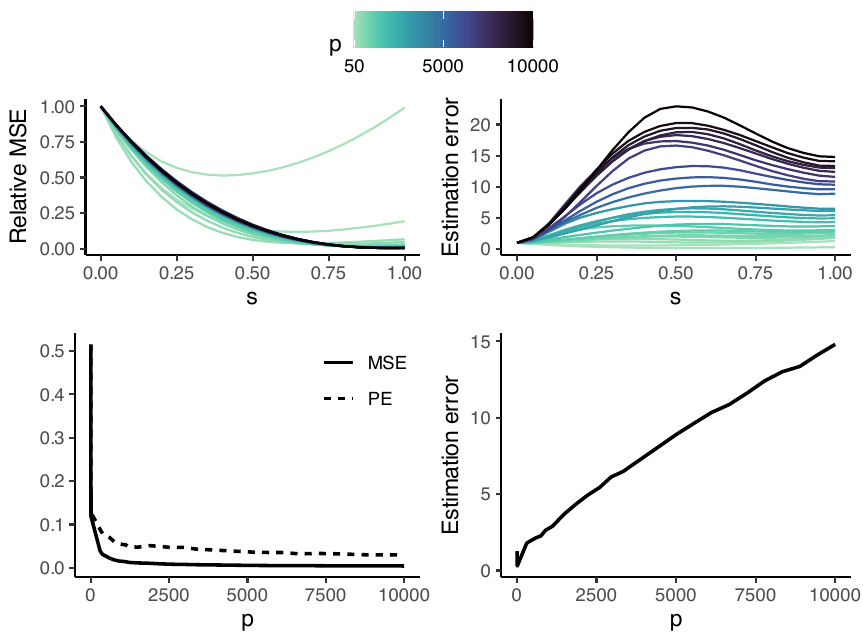}
\caption{$\Psi = \Psi_1 = I_p$. Top display: the relative MSE (left) and standardized $\ell_1$ estimation error (right) for $s \in \{0,...,1\}$ for different number of variables $p$. Bottom display: the relative MSE and relative PE at the optimal choice of $s$ (left), and the $\ell_1$ estimation error for $s = 0.5$ (right).}
\label{fig:fig2}
\end{figure}

\begin{figure}[H]
\centering
\includegraphics[width=0.9\linewidth]{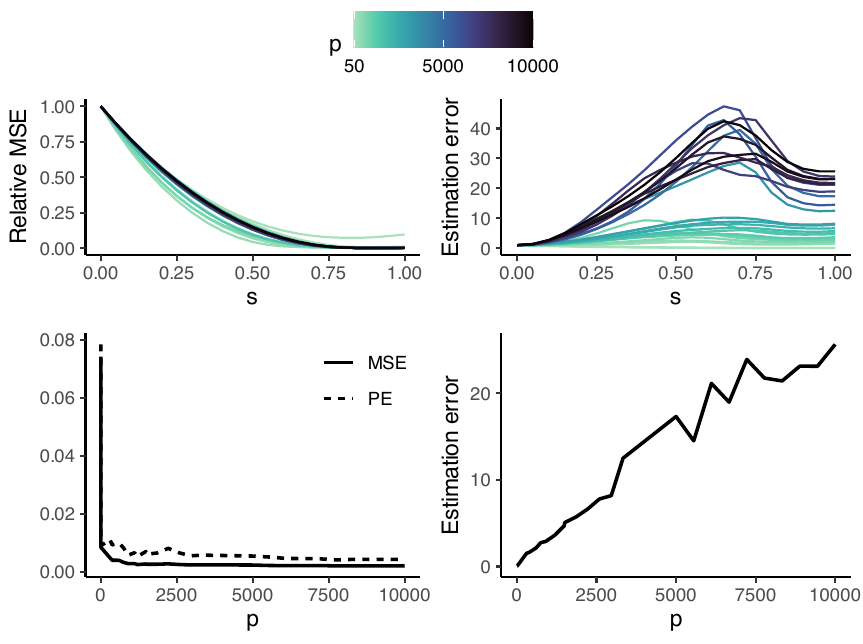}
\caption{$\Psi = \Psi_2$. Top display: the relative MSE (left) and standardized $\ell_1$ estimation error (right) for $s \in \{0,...,1\}$ for different number of variables $p$. Bottom display: the relative MSE and relative PE at the optimal choice of $s$ (left), and the $\ell_1$ estimation error for $s = 0.5$ (right).}
\label{fig:fig4}
\end{figure}

\begin{figure}[H]
\centering
\includegraphics[width=0.9\linewidth]{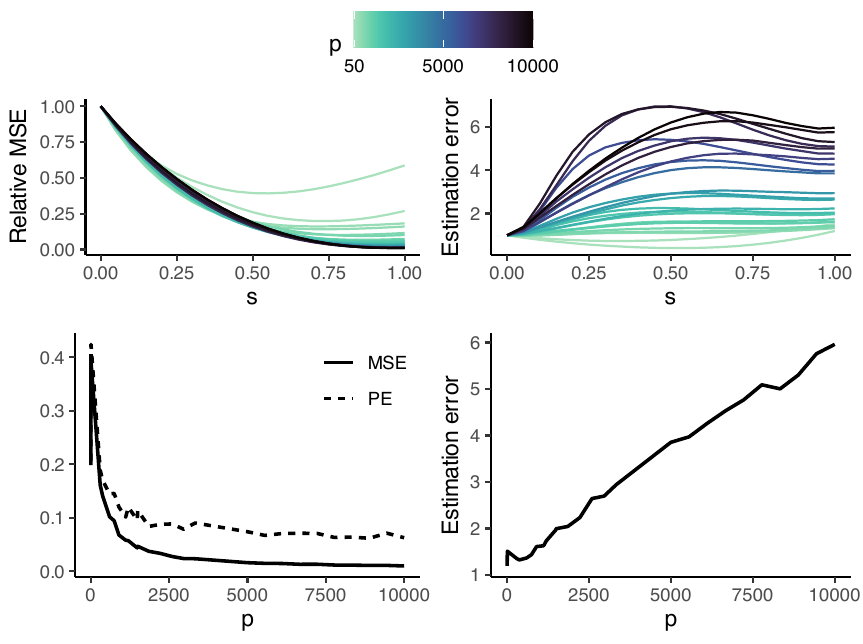}
\caption{$\Psi = \Psi_4$. Top display: the relative MSE (left) and standardized $\ell_1$ estimation error (right) for $s \in \{0,...,1\}$ for different number of variables $p$. Bottom display: the relative MSE and relative PE at the optimal choice of $s$ (left), and the $\ell_1$ estimation error for $s = 0.5$ (right).}
\label{fig:fig7}
\end{figure}

\begin{figure}[H]
\centering
\includegraphics[width=0.85\linewidth]{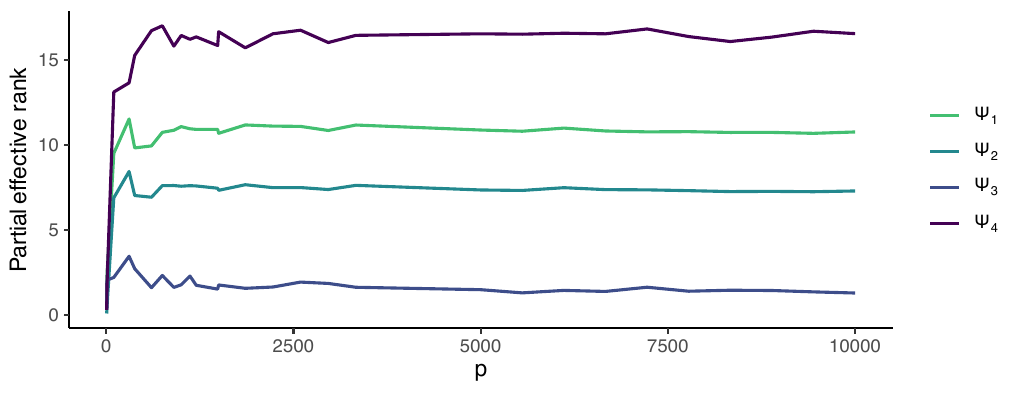}
\caption{The partial effective rank $\frac{1}{ \tilde{\omega}_1}\sum_{i = m+1}^{n} \tilde{\omega}_i$}
\label{fig:effectiverank}
\end{figure}

\newpage

\section{Impact of sequential removal of variables on Lasso prediction}
\label{sec:SequentialRemovalofVariables}

As briefly mentioned in Section \ref{sec:sparsityassumtionIntro}, \citet{ein2005outcome} showed that for one single breast cancer data set \cite{vanveer2002gene}, there exist several disjoint subsets of 70 genes that have almost the same predictive capabilities. From the simulation examples in Section \ref{sec:numericalstudies}, it is clear that under the latent variable model \eqref{EQ:latentvariablemodel}, similar conclusions can be drawn. To explore this, consider some general response vector $y$ and design matrix $X$, and a sequence of Lasso solutions $\{\hat{\boldsymbol{\gamma}}_1,...,\hat{\boldsymbol{\gamma}}_k\}$ where the initial estimate $\hat{\boldsymbol{\gamma}}_1$ is found by solving \eqref{EQ:penalizedformLasso} based on the entire design matrix $X$ and the optimal $\lambda$ is determined by cross-validation. The columns corresponding to the non-zero entries of $\hat{\boldsymbol{\gamma}}_1$ is then removed from the design matrix, and a subsequent Lasso estimate $\hat{\boldsymbol{\gamma}}_2$ is determined based on the remaining columns of X. In this way, the set selected in step 2 is disjoint from the first set. This process is repeated $k$ times, producing a sequence of selected subsets of variables that are all disjoint. Ideally, if the variables in the first set are substantially more important than the others, their removal would result in a much worse prediction performance of the subsequent models. We now explore the prediction performance of such sequences of Lasso estimates for a simulated example and for the real breast cancer data set of \citet{vanVijver2002gene}.

\subsection{Simulation example}
\label{subsec:sequentialremovalSimulation}

Consider the same artificial data as before, with $\Psi_1 = I_p$ and $p = 10000$. We compute a sequence of Lasso estimates of length $k = 5$, following the procedure outlined above. Table \ref{tab:table1} reports the mean $\text{MSE}(\hat{\boldsymbol{\gamma}}_k)$ and $\text{PE}(\hat{\boldsymbol{\gamma}}_k)$ over 100 replicates for $k \in \{1,...,5\}$. Observe that MSE and PE only gradually worsen with the removal of each "important" set. This relatively minor increase in prediction error suggests that, under the latent variable model, even when influential variables are removed, remaining variables can almost compensate and still give a relatively low prediction error. However, the fact that the prediction error does increase in this example is not surprising due to the diagonal $\Psi$; some variables are inevitably more predictive than others because they are measured with less noise. Table \ref{table:table3} shows the case with $\Psi = 0.5I_p$, which results in slower decay of performance across the active sets.

\begin{table}[!htb]
\centering
\begin{tabular}{llllll} \toprule
 &  \multicolumn{5}{c}{Active set no.}\\
& 1 & 2& 3 & 4 & 5 \\
\midrule
MSE & 0.046 & 0.050 & 0.054 & .059 & 0.061\\
PE & 0.152 & 0.165 & 0.192 & 0.240 & 0.279\\
\bottomrule
\end{tabular}
\caption{The in-sample and out-of-sample prediction errors for a sequence of cross-validated Lasso estimates.}
\label{tab:table1}
\end{table}

\begin{table}[!htb]
\centering
\begin{tabular}{llllll} \toprule
 &  \multicolumn{5}{c}{Active set no.}\\
& 1 & 2& 3 & 4 & 5 \\
\midrule
MSE & 0.028 & 0.026 & 0.026 & 0.029 & 0.033\\
PE &0.145 & 0.129 & 0.136 & 0.154 & 0.184\\
\bottomrule
\end{tabular}
\caption{The in-sample and out-of-sample prediction errors for a sequence of cross-validated Lasso estimates. The set-up is the same as in \ref{subsec:sequentialremovalSimulation} but with $\Psi = 0.5I$}
\label{table:table3}
\end{table}

\subsection{Real data example}
\label{sec:realdataexample}

Similarly to Section \ref{subsec:sequentialremovalSimulation}, we examine the effect of sequentially removing seemingly important variables in the real breast cancer data set of \cite{vanVijver2002gene}. This dataset originally consists of gene expression data from 24481 probes in 295 tumor samples. We use gene expression data to predict the risk of distant metastases, a common indicator of poor prognosis in cancer patients. The response is a binary variable, indicating whether or not the patient had distant metastases during the first five years after treatment. After filtering (i.e., removing probes with minimal or no variation and removing missing data), we are left with $4948$ genes and $289$ samples. We construct at a sequence of length $k=5$ of Lasso estimates and report the mean area under the curve (AUC) over 100 randomly generated training and test splits. For comparison, we perform the same analysis with Elastic net. Table \ref{tab:realdataset} shows the AUC for both the Lasso and Elastic net sequences. Notably, for Lasso, the AUC does not change much across step 1 and 2, which may suggest that there are at least two disjoint subset genes of comparable predictive accuracy. While the AUC scores are low (an AUC score of 0.5 corresponds to random guessing), only the changes in the AUC across the disjoint sets are of interest here. While the Elastic net solutions are consistently performing better than Lasso, this is mainly due to the size of the selected sets, which is considerably larger for Elastic net; the mean subset size is 859 and 26 for Elastic net and Lasso, respectively for the first active set.

\begin{table}[!htb]
\centering
\begin{tabular}{lllllll} \toprule
 &  \multicolumn{5}{c}{Active set no.}\\
& 1 & 2& 3 & 4 & 5 \\
\toprule
Lasso  & 0.654 & 0.675 & 0.593 & 0.521 & 0.516\\
Elastic net & 0.693 & 0.678 & 0.666 & 0.661  & 0.649 \\
\bottomrule
\end{tabular}
\caption{AUC scores for a sequence of Lasso and Elastic net estimates.}
\label{tab:realdataset}
\end{table}

\end{document}